\documentstyle[epsf,12pt]{article}
\newcommand{\beq}{\begin{equation}}
\newcommand{\eeq}{\end{equation}}
\newcommand{\beqa}{\begin{eqnarray}}
\newcommand{\eeqa}{\end{eqnarray}}

\newcommand{\bma}{\left( \begin{array}}
\newcommand{\ema}{\end{array} \right)}
\newcommand{\bfig}{\begin{figure}}
\newcommand{\efig}{\end{figure}}
\newcommand{\bpic}{\begin{picture}}
\newcommand{\epic}{\end{picture}}
\newcommand{\bc}{\begin{center}}
\newcommand{\ec}{\end{center}}
\newcommand{\pslash}{\kern 0.2 em p\kern -0.45em /}
\newcommand{\sla}[1]{\kern 0.2 em #1\kern -0.45em /}

\begin{document}
\setcounter{page}{0}
\thispagestyle{empty}
\hspace*{12.0cm}                        WU-B 93-35\\
\hspace*{12.0 cm}                     October 1993\\
\bc
{\Large\bf LARGE $p_{\perp}$ EXCLUSIVE PROCESSES: \\
           RECENT DEVELOPMENTS}\\
\vspace*{1.0 cm}
\ec
\bc
{\large P. Kroll}
\footnote{Supported in part by BMFT, FRG under contract 06 WU 737}
\footnote{E-mail: kroll@wpts0.physik.uni-wuppertal.de}
\footnote{Invited talk presented at
           the Conference on Hadron Structure 93, Banska Stiavnica (1993)}\\
\vspace*{0.5 cm}
Fachbereich Physik, Universit\"{a}t Wuppertal, Gau\ss strasse 20,\\
Postfach 10 01 27, D-5600 Wuppertal 1, Germany\\[0.3 cm]
\ec
\vspace*{4.0 cm}
\bc
                    {\bf Abstract}
\ec
Recent improvements of the hard scattering picture for exclusive
reactions, namely the inclusion of both Sudakov corrections
and the intrinsic transverse momentum dependence of the
hadronic wave function, are reviewed. On account of these improvements
the perturbative contribution to the pion's form factor can be calculated
in a theoretically self-consistent way for momentum transfers as low
about $2\,{\rm GeV}$. This is achieved at the expense of a substantial
suppression of the perturbative contribution in the few GeV region.
Eventual higher twist contributions are also discussed.\\
\newpage
\section{The hard scattering picture}
\setcounter{equation}{0}
\vspace*{-0.5cm}
There is general agreement that perturbative QCD in the framework of the
hard-scattering picture (HSP) \cite{lep:80} is the
correct description of form factors and perhaps other exclusive reactions
at asymptotically large momentum transfer. In the HSP a form factor
or a scattering amplitude is expressed by a convolution of distribution
amplitudes (DA) with hard scattering amplitudes calculated in collinear
approximation within perturbative QCD. The universal, process independent
DAs, which represent hadronic wave functions integrated over
transverse momenta ($k_{\perp}$), are controlled by long distance physics
in contrast to the hard scattering amplitudes which are governed by
short distance physics. The DAs cannot be calculated by perturbative
means, we have to rely on models. In principle lattice gauge theory
offers a possibility to calculate the DAs but with the present-day
computers a sufficient accuracy can not be achieved, only a few moments
of the pion and proton DAs have been obtained \cite{mar:88}.\\
As an example of an exclusive reaction let us consider the electromagnetic
form factor of the pion.
To lowest order pertubative QCD the hard scattering amplitude $T_H$ is
to be calculated from the two one-gluon exchange diagrams.
Working out the diagrams one finds
\beq
\label{eq:hs-amplitude}
T_H (x_1,y_1,Q,\vec k_\perp,\vec l_\perp) =
\frac{16 \pi \, \alpha_s(\mu) \, C_F}
{x_1 y_1 Q^2+(\vec k_\perp + \vec l_\perp)^2},
\eeq
where $Q (\geq 0)$ is the momentum transfer from the initial to the final
state pion. $x_1$ ($y_1$) is the longitudinal momentum fraction carried by
the quark and $\vec{k}_{\perp}$ ($\vec{l}_{\perp}$) its transverse momentum
with respect to the initial (final) state pion. The momentum of the
antiquark is characterized by $x_2=1-x_1$ ($y_2=1-y_1$) and
$-\vec{k}_{\perp}$ ($-\vec{l}_{\perp}$). $C_F$ ($=4/3$) is the colour
factor and $\alpha_s$ is the usual strong coupling constant to be evaluated
at a renormalization scale $\mu$. The expression (\ref{eq:hs-amplitude})
is an approximation in so far as only the most important $\vec{k}_{\perp}$-
and $\vec{l}_{\perp}$-dependences have been kept. Denoting the wave
function of the pion's valence Fock state by $\Psi_0$, the form factor
is given by
\beq
\label{eq:pre-hs-Fpi}
F_\pi(Q^2)=
\int \frac{dx_1 \, d^{\;\!2} k_\perp}{16 \pi^3}
\int \frac{dy_1 \, d^{\;\!2} l_\perp}{16 \pi^3} \,
\,\Psi_0^\ast (y_1,\vec l_\perp)\, T_H (x_1,y_1,Q,\vec k_\perp,\vec l_\perp)
\, \Psi_0 (x_1,\vec k_\perp).
\eeq
Strictly speaking $\Psi_0$ represents only the soft part of the pion
wave function, i.e. the full wave function with the perturbative tail
removed from it \cite{lep:80}. Contributions from higher Fock states are
neglected in (\ref{eq:pre-hs-Fpi}) since, at large momentum transfer, they
are suppressed by inverse powers of $Q^2$.\\
At large $Q$ one may neglect the $k_\perp$- and $l_\perp$-dependence
in the gluon propagator as well; $T_H$ can then be pulled out of
the transverse momentum integrals, and these integrations apply only to the
wave functions. Defining the DA as
\beq
\label{eq:DAdef}
\frac{f_\pi}{2 \sqrt{6}} \,\, \phi (x_1,\mu) =
\int \frac{d^{\;\!2} k_\perp}{16 \pi^3}\,  \Psi_0 (x_1,\vec k_\perp),
\eeq
one arrives at the celebrated hard-scattering formula for the pion's
form factor
\beq
\label{eq:hs-Fpi}
{F_\pi}^{HSP}(Q^2)=\frac{{f_\pi}^2}{24} \int dx_1 \, dy_1
\, \phi^\ast(y_1,\mu) \, T_H(x_1,y_1,Q,\mu) \, \phi(x_1,\mu),
\eeq
which is valid for $Q \to \infty$. The DA is defined such that
\beq
\label{eq:DAnorm}
\int_0^1 dx_1 \, \phi(x_1,\mu) = 1.
\eeq
An immediate consequence of the definitions (\ref{eq:DAdef}) and
(\ref{eq:DAnorm}) is that the constraint from the $\pi \to \mu\nu $ decay
\cite{lep:83} is automatically satisfied:
\beq
\label{eq:pi-munu}
\frac{f_\pi}{2 \sqrt{6}} = \int \frac{dx_1 \, d^{\;\!2} k_\perp}{16 \pi^3}
\,\Psi_0 (x_1,\vec k_\perp),
\eeq
where $f_\pi \,(=133\, {\rm MeV})$ is the usual $\pi$ decay constant. The
integral in (\ref{eq:DAdef}) has to be cut off at a scale of
order $Q$, which leads to a very mild dependence of the DA on the
renormalization scale (QCD evolution). An appropriate choice of the
renormalization scale is $\mu =\sqrt{x_1 y_1} Q$. This avoids large logs
from higher order pertubation theory at the expense, however, of a
singular behaviour of $\alpha_s$ in the end-point regions, $x_i, y_i \to 0$,
$i=1,2$. It is argued that radiative corrections (Sudakov factors) will
suppress that singularity and, therefore, in practical applications
of the HSP one may handle that difficulty by cutting off $\alpha_s$
at a certain value, which is typically chosen in the range of 0.5 to 0.7.
That crude receipe is unsatisfactory although the Sudakov argument itself is
correct as will be discussed in the next section.\\
Similarly to the pion's form factor other exclusive quantities can be
calculated at large transverse momentum. The HSP has two
characteristic properties, the quark counting rules
and the helicity sum rule. The first property says that the fixed
angle cross-section behaves at
large Mandelstam $s$ (and large transverse momentum) as
\beq
\label{a1}
\frac{d\sigma}{dt}= f (\theta) s^{2-n} \quad\quad (\rm{modulo\,\, logs})
\eeq
where n is the minimum number of external particles in the hard
scattering amplitude. The power laws also apply to form factors:
a baryon form factor behaves as $1/Q^4$, a meson form factor as $1/Q^2$
at large $Q$. These power laws are in surprisingly good agreement
with experimental data. Even at momentum transfers as low as 2 GeV
the data seem to respect the counting rules.\\
The second characteristic property of the
HSP is the conservation of hadronic helicity. For a two-body process,
$A B \to C D$ the helicity sum rule reads
\beq
\label{a2}
\lambda_A + \lambda_B = \lambda_C + \lambda_D.
\eeq
It appears as a consequence of utilizing the
collinear approximation and of dealing with (almost) massless quarks
which conserve their helicities when interacting with gluons.
The collinear approximation implies
that the relative orbital angular momentum between the constituents
has a zero component in the direction of the parent hadron. Hence the
constituents helicities sum up to the helicity of their parent hadron.
Experiments, e.g. the polarization in elastic proton-proton scattering
\cite{cam:85} or the recent measurement ot the proton's Pauli
form factor \cite{bos}, reveal that the hadronic helicity is not conserved;
the ratio of hadronic helicity flip matrix elements to non-flip ones is
about 0.2 - 0.3. That fact can be regarded as a hint at sizeable higher twist
contributions in the few GeV region. The physical origin of these
higher twist contributions is not yet known. The higher twist
nature of the Pauli form factor is clearly visible in
Fig.~1: $F_2$ behaves as $1/Q^6$ at large $Q$. It is important
to realize that the HSP in its present form cannot predict such higher
twist terms.\\
\bfig[t]
\bc
\unitlength 1cm
\bpic(9,6)
\epsfbox{f2.psc}
\epic
\ec
\caption[abcd]{\it
The Pauli form factor of the proton scaled by $Q^6$. Data are taken
from Ref.~{\rm\cite{bos}}. The solid line represents the results obtained
from the diquark model {\rm\cite{jkss:93}}.}
\efig
There are many calculations of large $p_{\perp}$ exclusive reactions
within the framework of the HSP: Electromagnetic form factors
of mesons and baryons, $N \to N^*$ transition form factors, Compton
scattering off nucleons, photoproduction of mesons, two-photon
annihilations into pairs of mesons or baryons, decays of heavy
mesons such as $\psi \to p\bar p$ or $B\to \pi\pi$. No clear picture
has been emerged as yet; there are successes and failures. It however
seems that one only obtains results of the order of the experimental
values if, at least for the proton and the pion, DAs are used which are
strongly concentrated in the end-point regions. Such DAs, first
proposed by Chernyak and Zhitnitsky (CZ) \cite{CZ:82}, find a certain
justification in QCD sum rules by means of which a few moments of
the DAs have been calculated. The CZ moments are subject of considerable
controversy: Other QCD sum rule studies provide other values for
the moments \cite{rad:91}. Likewise the results obtained from lattice
gauge theories do not well agree with the CZ moments \cite{mar:88}.\\
On the other hand, the asymptotic forms of the DAs ($\sim x_1 x_2$ for
the pion, $\sim x_1 x_2 x_3$ for the proton), into which any DA evolves
for $Q\to\infty$, lead to results which are typically orders of
magnitudes too small as compared with data. Consider, as an example,
the magnetic form factor of the proton. For the DAs of the CZ type one obtains
$Q^4 G_M \sim 1 \rm{GeV}^4$ in agreement with experiment, whereas
a vanishing result is found for the asymptotic DA.\\
Purely hadronic reactions, as for instance elastic proton-proton
scattering, have not yet been studied in the frame work of the HSP.
The reason for that fact is, on the one hand, the huge number of
Feynman diagrams contributing to such reactions and, on the other hand,
the occurence of multiple scatterings (pinch singularities \cite{lan:74}),
i.e. the possibility that pairs of constituents scatter independently
in contrast to the HSP in which all constituents collide within a small
region of space-time. A general framework for treating multiple
scattering contributions has been developed by Botts and
Sterman \cite{bot:89}.
\vspace*{-0.5cm}
\section{The Botts-Li-Sterman approach}
\setcounter{equation}{0}
\vspace*{-0.5cm}
The applicability of the HSP at experimentally accessible momentum
transfers, typically a few GeV, was questioned \cite{rad:91,Isg:89}. It
was asserted that in the few GeV region the hard-scattering picture
accumulates large contributions from the soft end-point regions, rendering
the perturbation calculation inconsistent. This is in particular the case
for the end-point concentrated DAs. Another source of theoretical
inconsistency is caused by the collinear approximation: The neglect of the
transverse momentum dependence of the hard scattering amplitude,
see for instance eq.~(\ref{eq:hs-amplitude}), is a bad approximation in
the end-point regions. How strongly the results, say for the pion's form
factor (\ref{eq:hs-Fpi}), are distorted by that approximation depends
on the shape of the DAs. Obviously, for DAs of the CZ type, for which
the end-point regions get strong weights, the neglect of the transverse
momentum dependence of the hard scattering amplitude entails
large errors in the final results. Therefore, that approximation cannot
be retained for the DAs of the CZ type. For the asymptotic DA and similar
forms the approxiations turns out to be reasonable. For details see Sect.~3.\\
The statements made by the authors of \cite{rad:91,Isg:89} were
challenged by Sterman and collaborators \cite{bot:89,LiS:92,Li:92}.
These authors suggest to retain the transverse momentum dependence
of the hard scattering amplitude and to take into account Sudakov
corrections, suppressing the dangerous end-point regions even further.
In order to include the Sudakov corrections it is advantageous to
reexpress eq.~(\ref{eq:pre-hs-Fpi}) in terms of the Fourier transform
variable $\vec b$ in the transverse configuration space
\beqa
\label{eq:ft-Fpi}
{F_\pi}^{pert}(Q^2)&=&
\int_0^1\! \frac{dx_1 \, dy_1}{(4 \pi)^2}
\int_{-\infty}^\infty \!d^{\;\!2}b
\,\hat{\Psi}_0^\ast (y_1,\vec b,w)\,\hat{T}_H (x_1,y_1,Q,b,t)
\,\hat{\Psi}_0 (x_1,-\vec b,w) \nonumber \\
& &\qquad\qquad\qquad\qquad\qquad
\times\,\exp\left[-S(x_1,y_1,Q,b,t)\right]
\eeqa
where the Fourier transform of a function $f=f(\vec k_\perp)$ is
denoted by $\hat{f}=\hat{f}(\vec b)$. As the renormalization scale
Sterman et al. choose the largest mass scale appearing in $\hat{T}_H$:
\beq
\label{eq:als-arg}
t= {\rm Max}(\sqrt{x_1 y_1}\,Q,w=1/b).
\eeq
The factor $\exp \left[-S\right]$ in
(\ref{eq:ft-Fpi}), termed the Sudakov factor, comprises the radiative
corrections. The lengthy expression for the Sudakov
exponent $S$, which includes all leading and next-to-leading
logarithms, is given explicitly in \cite{LiS:92}. The most important
term in it is the double logarithm
\beq
\label{eq:double-log}
\frac{2}{3\beta_1} \ln \frac{\xi Q}{\sqrt{2} \Lambda_{QCD}}
\ln \frac{\ln (\xi Q/\sqrt{2} \Lambda_{QCD})}{\ln (1/b \Lambda_{QCD})},
\eeq
where $\xi$ is one of the fractions, $x_i$ or $y_i$, and
$\beta_1= (33-2n_f)/12$. For small $b$, i.e. at small transverse
separation of quark and antiquark, there is no suppression from the
Sudakov factor. As $b$ increases the Sudakov factor decreases,
reaching zero at $b=1/\Lambda_{QCD}$ (see Fig.~2). For $b$ larger
than $1/\Lambda_{QCD}$ the Sudakov factor is set to zero. Owing to
that cut-off the singularity of $\alpha_s$ is avoided without
introducing a phenomenological parameter (e.g.~a gluon mass). For
$Q \to \infty$ the Sudakov factor damps any contribution except
those from configurations with small quark-antiquark separation.
In other words, the hard-scattering contributions dominate the pion's
form factor asymptotically.\\
\bfig[t]
\bc
\unitlength 1cm
\bpic(7.5,4.5)
\put(0,0.5){\epsfbox{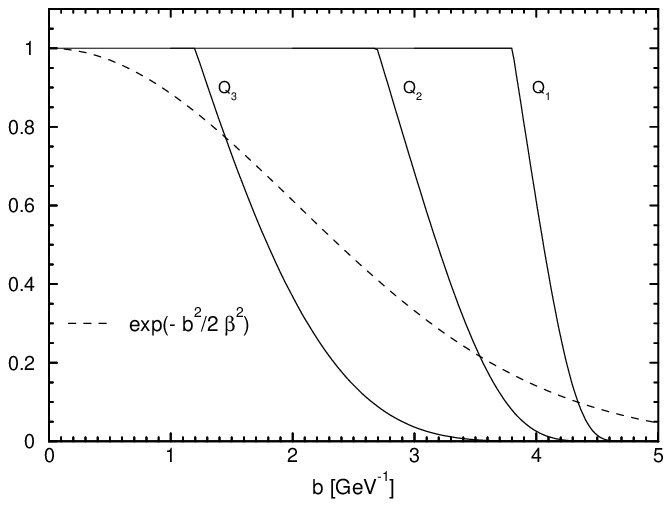}}
\epic
\ec
\caption[bcde]{\it
The Sudakov factor, evaluated at $x_1=y_1=1/2$, and the Gaussian
$\exp (-b^2/2\beta^2)$ (see eq.~{\rm (\ref{eq:ft-gaussian})}) as functions of
the transverse separation $b$. The Gaussian is shown for a r.m.s. transverse
momentum of $350\,{\rm MeV}$ (dashed line). The Sudakov factor is evaluated
at $Q_1=2\,{\rm GeV}$, $Q_2=5\,{\rm GeV}$ and $Q_3=20\,{\rm GeV}$ with
$\Lambda_{QCD}=200\,{\rm MeV}$ (solid lines).}
\efig
Li and Sterman have explored the improved hard-scattering
formula (\ref{eq:ft-Fpi}) on the basis of customary
wave functions, neglecting the QCD evolution and the intrinsic transverse
momentum dependence of the wave functions. Their numerical studies
have revealed
that the modified perturbative approach is self-consistent for
$Q>20\Lambda_{QCD}$ in the sense that less than, say, $50\%$ of the
result is generated by soft gluon exchange $(\alpha_s > 0.7)$. In
the few GeV region the values for $F_\pi$ as obtained by Li and Sterman
are somewhat smaller than those provided by the hard-scattering formula
(\ref{eq:hs-Fpi}) and are, perhaps, smaller than the experimental
values \cite{Beb:76}, see Fig.~3.
\bfig[t]
\bc
\unitlength 1cm
\bpic(16,5)
\put(0,1.5){\epsfbox{fig2a.psc}}
\hspace*{1.0cm}
\put(8,1.5){\epsfbox{fig2b.psc}}
\epic
\ec
\caption[figname]{\it
(Left) The pion's form factor as a function of $Q^2$ evaluated
with the CZ wave function and $\Lambda_{QCD}=200\,{\rm MeV}$. The dash-dotted
line is obtained from the hard-scattering formula with $\alpha_s$ cut off
at 0.5 and the dashed line from {\rm(\ref{eq:ft-Fpi})} ignoring the intrinsic
$k_\perp$-dependence. The solid line represents the complete result obtained
from {\rm(\ref{eq:ft-Fpi})} taking into account both the Sudakov factor and the
intrinsic $k_\perp$-dependence
($\langle k_\perp^2 \rangle^{1/2} =350\,{\rm MeV}$).
Data are taken from {\rm\cite{Beb:76}} ($\circ$ 1976, $\bullet$ 1978).\\
(Right)As left figure but using the MAS wave function. Note the modified
scale of the abscissa.}
\efig
\vspace*{-0.5cm}
\section{The intrinsic $k_\perp$-dependence of the hadronic wave function}
\setcounter{equation}{0}
\vspace*{-0.5cm}
The approach proposed by Sterman and collaborators \cite{bot:89,LiS:92,Li:92}
certainly constitutes an enormous progress in our understanding of
exclusive reactions at large momentum transfer.
In any practical application of that approach one has however to allow
for an intrinsic transverse momentum dependence of the hadronic wave
function \cite{jak:93}, although, admittedly, this requires a new
phenomenological element in the calculation. Fortunately, for the case of
the pion the intrinsic transverse momentum of its valence Fock
state wave function is well constrained.\\
In accordance with (\ref{eq:DAdef}), (\ref{eq:DAnorm}) and
(\ref{eq:pi-munu}), the wave function can be
written as
\beq
\label{eq:wvfct}
\Psi_0 (x_1,\vec k_\perp) = \frac{f_\pi}{2 \sqrt{6}}
\,\phi(x_1) \,\Sigma(x_1,\vec k_\perp),
\eeq
the function $\Sigma$ being normalized in such a way that
\beq
\label{eq:Sigmanorm}
\int \frac{d^{\;\!2} k_\perp}{16 \pi^3} \,\Sigma (x_1,\vec k_\perp) =1.
\eeq
The wave function (\ref{eq:wvfct}) is subject to the following
constraints: It is normalized to a number $P_{q\bar{q}}\leq 1$, the
probability of the valence quark Fock state; the value of the
configuration space wave function at the origin is determined by the
the $\pi$ decay constant (see eq.~(\ref{eq:pi-munu})); the process
$\pi ^0 \to \gamma\gamma$ provides a third relation \cite{lep:83}.
Finally, the charge radius of the pion provides a lower limit on the
root mean square (r.m.s.) transverse momentum, actually it
should be larger than $300 \,{\rm MeV}$.
The $k_\perp$-dependence of the wave function is parametrized as
a simple Gaussian
\beq
\label{eq:Sigma}
\Sigma(x_1,\vec k_\perp)=16 \pi^2 \beta^2 \,g(x_1)
\,\exp \left(-g(x_1) \beta^2 k_\perp^2 \right),
\eeq
$g(x_1)$ being either  $1$ or $1/x_1 x_2$. The latter case goes along
with a factor $\exp (-\beta^2 m_q^2/x_1 x_2 ) $ in the DA
where $m_q$ is a constituent quark mass ($330 \,{\rm MeV}$). The Gaussian
(\ref{eq:Sigma}) is consistent with the required large-$k_\perp$ behaviour
of a soft wave function. Several wave functions have been employed in
\cite{jak:93}. Here, in the present paper, only the results for the two
extreme cases utilized in \cite{jak:93} are quoted. That is, on the one hand,
the CZ wave function $\sim x_1 x_2 (x_1-x_2)^2$ , $g=1$ \cite{CZ:82}
which is the example most concentrated in the end-point region and,
on the other hand, the modified asymptotic (MAS) wave function
$\sim x_1 x_2$, $g=1/x_1 x_2$ \cite{lep:83}. The MAS wave function is the
example least concentrated in the end-point regions.\\
The wave functions have one free parameter, the oscillator
parameter $\beta$, which is fixed by requiring specific values for the
r.m.s. transverse momentum. For a value of $350\,{\rm MeV}$ all the
constraints on the pion wave functions are well respected \cite{jak:93}.
For a value of $250\,{\rm MeV}$ for instance the constraint
from $\pi^0\to\gamma\gamma$ decay is badly violated and the radius of the
pion is too large \cite{jak:93}.\\
The Fourier transform of the $k_\perp$-dependent part of
the wave function reads
\beq
\label{eq:ft-gaussian}
\hat \Sigma (x_1,\vec b)=
4\pi\,\exp\left(-\frac{b^2}{4 g(x_1) \beta^2}\right).
\eeq
Li and Sterman \cite{LiS:92} assume that the dominant $b$-dependence
of the integrand in eq.~(\ref{eq:ft-Fpi}) arises from the Sudakov
factor and that the Gaussian in $\hat\Sigma$ can consequently be
replaced by 1. In order to examine that assumption,
the Gaussian is compared with the Sudakov factor in Fig.~2. Obviously the
intrinsic  $k_\perp$-dependence of the wave function cannot be
ignored. For momentum transfers of the order of a few GeV the
wave function damps the integrand in (\ref{eq:ft-Fpi}) more than
the Sudakov factor. Only at very large values of $Q$ does the
Sudakov factor take over.\\
Numerical evaluations of the pion's form factor
through (\ref{eq:ft-Fpi}), using the various wave functions mentioned
above, confirm the observations made in Fig.~2. The intrinsic
transverse momentum dependence of the wave function provides additional
suppression to that due to the Sudakov factor (see Fig.~3). The
suppression is particular strong for the end-point concentrated
wave functions. These observations confirm the statements made
at the end of Sect.~1: The so-called success of the DAs of the CZ-type
is only fictitious; for finite values of $Q$ the HSP formula
(\ref{eq:hs-Fpi}) does not represent a reasonable approximation
to (\ref{eq:pre-hs-Fpi}) for such DAs. The neglect of the
$k_\perp$-dependence in the hard scattering amplitude is unjustified
in that case.\\
Thus one can conclude that the intrinsic $k_\perp$-dependence of
the wave function has to be taken into account for a reliable
quantitative estimate of the perturbative QCD contribution to the pion's
form factor. One has to be aware that this introduces a new phenomenological
element into the calculation. That disadvantage is, at least
partially, compensated by the fact that the inclusion of the
intrinsic $k_\perp$-dependence renders the perturbative contribution
even more self-consistent than the Sudakov suppression already
does. Applying the criterion of self-consistency as suggested
by Li and Sterman \cite{LiS:92} (see above), we can conclude that
perturbative QCD begins to be self-consistent for $Q$ at about
$2\,{\rm GeV}$ (for $\langle k_\perp^2 \rangle^{1/2}=350\,{\rm MeV}$).
The exact value of $Q$ at which self-consistency sets in depends on the wave
function. It is larger for the end-point concentrated wave
functions than for the asymptotic or MAS DAs.
However, the perturbative
contribution (\ref{eq:ft-Fpi}), although self-consistent, is
presumably too small with respect to the data. A definite conclusion on
that point is unfortunately not yet possible since the data may suffer
from large systematic errors \cite{Car:90}.\\
An analogous investigation of the proton's form factor, for which precise
data is at our disposal, may allow a decisive conclusion on the agreement
between the perturbative contribution and experiment. Such an analysis
has unfortunately not yet been carried out.\\
Suppose the contributions from the improved HSP are indeed
too small for the pion's and the proton's form factor. Hence
other contributions must play an important role
in the few GeV region. Obviously, for a perturbative calculation one may
suspect higher order contributions to be responsible for the discrepancy
between theory and experiment. In analogy to the Drell-Yan process
such contributions may be condensed in a K-factor multiplying the
lowest order result for the form factor
\beq
\label{kfac}
K = 1 + \frac{\alpha_S (\mu)}{\pi} B(Q, \mu) + \cal O (\alpha_S).
\eeq
Calculations of the one-loop corrections \cite{fie:81,dit:81} reveal
that the magnitude of the $K$-factor strongly depends
on the renormalization scale. It is in general large except the renormalization
scale is chosen like $\mu = \sqrt{x_1 y_1} Q$ (see Sect.~1). With such a choice
and the use of the asymptotic DA, K is about 1.3 in the few GeV region.
For DAs broader than the asymptotic one, i.e.~for such with a stronger
weight of the end-point regions, B seems to become negative. Note that at
least part of the K-factor is included in the Sudakov factor. With regard
to the new developments discussed above it is perhaps advisable to reanalyse
the $\alpha_S^2$ corrections.\\
Missing soft contributions offer another explanation of the eventual
discrepancy between theory and experiment. As the $k_\perp$-effects
discussed above such contributions are of higher twist type and do not
respect the quark counting rules. Dominance of such contributions in the
case of the pion's form factor and perhaps in other exclusive quantities
would leave unexplained the apparent agreement between the counting rules
and experiment (see Sect.~1).\\
There are several possible sources for such soft contributions:\\
i) Genuine soft contributions like VMD contributions or contributions
from the overlap of the soft parts of the hadronic wave functions
may fill in the eventual gap between the pertubative contribution and
experiment. Allowing for sufficiently many vector mesons the VMD models are
flexible enough to account for the form factor data even at large $Q$
(see for instance Ref.~\cite{dub:93}). The overlap contribution can be
estimated with the aid of the famous Drell-Yan formula \cite{Dre:70} (note
that the HSP represents the contribution from the overlap of the perturbative
tails of the hadronic wave functions). Using the wave functions
(\ref{eq:wvfct}), (\ref{eq:Sigma}) one finds for the pion case
\beq
\label{eq:Fpi-soft}
{F_\pi}^{soft}(Q^2)= \frac{\pi^2}{3}\,{f_\pi}^2\,\beta^2
\int dx_1 \,g(x_1)\,\phi^2(x_1)\,
\exp\left( -g(x_1)\,\beta^2\, {x_2}^2\, Q^2/2 \right).
\eeq
The integral is dominated by the region near $x_1=1$, other regions are
strongly damped by the Gaussian. For example, taking
$g=1$, the effective region is $1-\sqrt{2}/Q\beta \geq x_1 \geq 1$.
Hence $F_\pi^{soft}$ sensitively reacts to the behaviour of the DA for
$x_1 \to 1$. Evaluations of (\ref{eq:Fpi-soft}) reveal that the MAS wave
function provides a soft contribution of the right magnitude to fill
in the gap between the perturbative contribution (\ref{eq:ft-Fpi}) and the
experimental data (see Fig.~4). As required by
the consistency of the entire picture, $F_\pi^{soft}$
decreases faster with increasing $Q$ than the perturbative contribution.
The exponential $\exp(-\beta^2 m_q^2/x_1 x_2)$ multiplying the asymptotic DA
turns out to be quite important since it is effective in the end-point
regions: It reduces the size of $F_\pi^{soft}$ substantially and leads
to an exponentially damped decrease for $Q\to\infty$. $F_\pi^{soft}$ becomes
equal to the perturbative contribution at $Q\simeq 5\,{\rm GeV}$. The
soft contributions are also subject to Sudakov corrections. For the MAS wave
function these corrections amount to a few per cent.
The CZ wave function provides a very large soft contribution because of
its strong concentration in the end-point regions. The size of that
contribution is extremely sensitive to details such as the QCD evolution.
Therefore, this wave function appears to be unrealistic. Soft
contributions of the type (\ref{eq:Fpi-soft}) have also been discussed
in \cite{Isg:89}. Our results for $F_\pi^{soft}$ are similar in trend, but
smaller in size than those presented in \cite{Isg:89}.
Strong soft contributions to form factors have also been obtained with
QCD sum rules \cite{rad:91}.\\
\bfig[t]
\bc
\unitlength 1cm
\bpic(9,6)
\put(0,0){\epsfbox{vgl_hua.psc}}
\epic
\ec
\caption[figdummy]{\it
Comparison of the soft (dashed line) and the perturbative (dotted line)
contributions to the pion's form factor evaluated for the MAS
wave function.}
\efig
I would like to emphasize that in the HSP as well as in the Drell-Yan
formula the soft hadronic wave function is input. It is an unknown
function since it cannot be calculated
with a sufficient degree of accuracy from QCD at present.
Thus, to some extend, a study of a form factor serves rather
as a determination of the wave function than as a prediction of the
form factor. A predictive power is only achieved if that wave function
can be used to calculate other reactions. In the HSP this is possible:
For instance, the DA of the proton is fixed in a study of the form factor
and subsequently used in order to predict, say, Compton scattering off
protons. In general soft models do not allow the study of other reactions
without introducing new unknown parameters and/or functions. \\
ii) There may be orbital angular momentum components in the hadronic
wave function other than zero. For instance, the valence Fock state
component of the pion may be expressed by
\beq
\label{ang}
\int\frac{dx_1\,d^{\;\!2}k_\perp}{16\pi^3}
\frac{1}{\sqrt{2}}\, \left(\sla{p}  +m \right)\,
\left[\Psi^0(x_1,\vec k_\perp)
+ \sla{k}_\perp \Psi^1(x_1,\vec k_\perp)\right]\, \gamma_5 ,
\eeq
where $p$ denotes the pion's momentum and $m$ its mass. A new
phenomenological function, $\Psi^1$ appears in (\ref{ang}) which is certainly a
disadvantage but may lead to a better quantitative
description of the pion's form factor data. Treating quark and antiquark
as free particles, the expansion of their spinors around the direction
of their parent pion provides a model function for $\Psi^1$:
\beq
\label{psi1}
\Psi^1 = \Psi^0/x_1 x_2.
\eeq
$L \neq 0$ components may also appear in other hadronic wave functions.
They have the appealing consequence that the helicity sum rule is violated
for finite values of $Q$. This may offer the possibility
to calculate the Pauli form factor of the proton.\\
iii) Contributions from higher Fock states are another source of higher twist
contributions.
Also in this case new phenomenological functions have to be introduced.\\
iv) For baryons one may also think of quark-quark correlations in the wave
functions which also constitute higher twist effects. In a series of papers
\cite{kro:93,pil:93,jkss:93} the idea has been put forward that such
correlations can effectively be described as quasi-elementary diquarks.
A systematic study of all exclusive photon-proton reactions has been
carried out in that diquark model, which is a variant of the unmodified HSP:
form factors in the space-like and in the time-like regions, virtual and
real Compton scattering, two-photon-annihilations into proton-antiproton
as well as photoproduction of mesons.
A fair description of all the data has been achieved utilizing in all
cases the same proton DA (as well as the same values for the
few other parameters specifying the diquarks). The diquark model
allows to calculate helicity flip amplitudes and consequently to predict
for instance the Pauli form factor of the proton. The results for it, shown in
Fig.~1, are in agreement with the data. Results for the magnetic form
factor in both the time-like and the space-like regions are shown in Fig.~5.
Although the proton DA used in these studies is not strongly concentrated
in the end-point regions, it is rather of the MAS type,
it still remains to be seen how much Sudakov factors and the intrinsic
$k_\perp$-dependence of the proton wave function will change the results.
\bfig[t]
\bc
\unitlength 1cm
\bpic(9,6)
\put(0,0){\epsfbox{proton.psc}}
\epic
\ec
\caption[figdummy]{\it
The magnetic form factor of the proton in the time-like and space-like
(at $Q^2=-s$) regions. The time-like data are taken from
Ref.~{\rm\cite{arm:93}}, the space-like data from Ref.~{\rm\cite{arn:86}}.
The solid lines represent the predictions of the diquark model
{\rm\cite{pil:93}}, the dashed line is Hyer's prediction {\rm\cite{hy:93}}.}
\efig
\vspace*{-0.5cm}
\section{Other applications of the modified HSP}
\setcounter{equation}{0}
\vspace*{-0.5cm}
Up to now only a few applications of the modified HSP have been
published. More work is urgently needed. According to what I said above
a systematic study of all the photon-proton reactions in that framework
would be extremely important. From a comparison between predictions
and the many accurate data we have at our disposal for that class
of reactions one may be able to draw definite conclusions about the
magnitude of the higher twist contributions and perhaps to elucidate their
nature.\\
Li \cite{Li:92} has calculated the magnetic form factor of the proton or
rather the Dirac form factor $F_1$ in the space-like region. Fair
agreement with the data is obtained if a DA of the CZ type is used. The
calculation turns out to be self-consistent for $Q \geq 30\Lambda_{QCD}$.
With a value of $200\,\rm{MeV}$ for $\Lambda_{QCD}$ the region of
self-consistency is beyond the measured region. The intrinsic
$k_\perp$-dependence of the wave function has not been taken into account
by Li. I expect, on the basis of our experience with the pion's form
factor, that the region of self-consistency is extendend down to much
smaller values of $Q$ if the intrinsic $k_\perp$-dependence is taken
into account but this will likely be achieved at the expense
of a suppression of the perturbative contribution.\\
Coriano et al.~\cite{cor:92} have calculated the academic process
Compton scattering off pions in that framework. Again the intrinsic
$k_\perp$-dependence of the pion wave function has not been taken
into account.\\
Finally, Hyer \cite{hy:93} has investigated the magnetic form factor
of the proton in the time-like region as well as $\gamma\gamma\to p \bar p$.
According to Hyer the latter process can only be calculated in a
self-consistent way at rather large values of $s$ which prevents a
comparison with the new data from CLEO \cite{ong:92}. The predictions
from the diquark model, on the other hand, agree very well with the CLEO
data \cite{pil:93}. For the time-like form factor of the proton Hyer
finds values which are about a factor of 1.5 larger than those obtained
by Li \cite{Li:92} for space-like form factor. The reason for Hyer's
result is not clear: He has not analytically continued the Sudakov
factor and the gluon propagators. Therefore, one would expect from this
calculation about the same
values for the form factor in both the regions the time-like and space-like
ones. Comparing with the data of the Fermilab E760 collaboration
\cite{arm:93} (see Fig.~5), Hyer's result is still somewhat too small.
Thus, one is tempted to consider the large difference between the time-like
and space-like data for the magnetic form factor of the proton as another
hint at strong soft contributions in the $5 - 15\, \rm{GeV}^2$ region. Note
that
the diquark model accounts quite well for that difference.
\vspace*{-0.5cm}
\section{Summary}
\vspace*{-0.5cm}
The improved HSP which includes both the Sudakov corrections and the intrinsic
$k_\perp$-dependence of the hadronic wave function constitutes an
enormous progress in our understanding of exclusive reactions although
there are still some theoretical problems left over. At
least for the pion's form factor it has been shown that the improved HSP
allows to calculate the perturbative contribution to that form factor
in a theoretically self-consistent way for momentum transfers as low
as about $2\,{\rm GeV}$. This is, however, achieved at the expense of a
strong suppression of the perturbative contribution as compared to that
obtained with the original unmodified HSP. Now the perturbative contribution
is likely too small as compared to data. Similar results are to be
expected for other exclusive reactions, as for instance the magnetic
form factor of the proton. It thus seems that other contributions
(higher twists) also play an important role in the few GeV region as already
indicated by polarization effects observed in various exclusive
reactions and by the recently measured Pauli form factor of the proton.
An interesting task for the future is to find out the size of such
higher twist contributions and to elucidate their physical nature.
Finally, I would like to emphasize that the approximate validity of the quark
counting rules, for which the HSP offers an explanation, would remain
a mystery if all data for large $p_\perp$ exclusive processes are
dominated by soft contributions.\\
Acknowledgements: I would like to thank S.~Dubnicka and A.~Z.~Dubnickova
for the enjoyable and stimulating atmosphere of this conference.

\end{document}